\documentclass[]{spie}  

\usepackage[]{graphicx}

\title{Telluric-line subtraction in high-accuracy velocimetry: a PCA-based approach} 

\author{\'Etienne Artigau\supit{a},  Nicola Astudillo-Defru\supit{b}, Xavier Delfosse\supit{b},  Fran\c cois Bouchy\supit{c}, Xavier Bonfils\supit{c}, Christophe Lovis\supit{d}, Francesco Pepe\supit{d},  Claire Moutou\supit{e}, Jean-Fran\c cois Donati\supit{f}, Ren\'e Doyon\supit{a} and Lison Malo\supit{e}
\skiplinehalf
\supit{a}D\'epartement de physique and Observatoire du Mont-M\'egantic, Universit\'e de Montr\'eal, Montr\'eal H3C 3J7, Canada; \\
\supit{b}Institut de Plan\'eologie et d'Astrophysique de Grenoble, UMR 5274 CNRS, Universit\'e Joseph Fourier, BP 53, 38041 Grenoble Cedex 9, France;\\
\supit{c}Aix Marseille Universit\'e, CNRS, LAM (Laboratoire d'Astrophysique de Marseille) UMR 7326, 13388 Marseille, France;\\
\supit{d}Observatoire Astronomique de l'Universit\'e de Gen\`eve, 51 Ch. des Maillettes, 1290 Sauverny, Versoix, Switzerland; ;\\
\supit{e}CFHT Corporation, 65-1238 Mamalahoa Hwy Kamuela, Hawaii 96743, USA;\\
\supit{f}IRAP-UMR 5277, CNR and Universit\'e de Toulouse, 14 Av. E. Belin, F-31400 Toulouse, France
}

\authorinfo{Further author information: (Send correspondence to E.A., artigau@astro.umontreal.ca, Telephone: 1 514-343-6111, ext. 3190}

\begin{document} 
  \maketitle 

\begin{abstract}
Optical velocimetry has led to the detection of more than 500 planets to date and there is a strong effort to push m/s velocimetry to the near-infrared to access cooler and lighter stars. The presence of numerous telluric absorption lines in the nIR brings an important challenge. As the star's barycentric velocity varies through the year, the telluric absorption lines effectively varies in velocity relative to the star's spectrum by the same amount leading to important systematic RV offsets. We present a novel Principal component analysis-based approach for telluric line subtraction and demonstrated its effectiveness with archival HARPS data for GJ436 and $\tau$ Ceti, over parts of the $R$-band that contain strong telluric absorption lines. The main results are: 1) a better RV accuracy with excluding only a few percentage of the domain, 2) better use of the entire spectrum to measure RV and 3) a higher telescope time efficency by using A0V telluric standard from telescope archive.
\end{abstract}



\section{INTRODUCTION}
\label{sec:intro}  

The presence of strong telluric absorption lines in the near infrared represents a challenge for m/s velocimetry as it adds a spectral component to all spectra that can be offset by as much as $30$\,km/s  relative to the stellar lines. While the optical domain presents relatively large wavelength intervals free of telluric features (say above a few \%), most of the near-infrared domain is affected. Proper handling of telluric absorption would therefore lead to a significant increase in the domain containing Doppler information.  The present work has been performed in the context of the development of the SPIRou spectropolarimeter\cite{Artigau:2011, Delfosse:2013}, but is relevant for stable, high-resolution ($\lambda/\delta\lambda>$20\,000), near-infrared spectrograph such as CARMENES\cite{Quirrenbach:2010gf}, GIANO\cite{Oliva:2012} or IRD\cite{Tamura:2012}.

\section{PCA approach to telluric absorption}
We propose a novel approach to telluric line absorption. This approach is based on a simple physical assumption; the observed telluric absorption spectrum is the sum (in absorbance) of a finite (H$_2$O, O$_2$, CO$_2$, N$_2$O, CH$_4$) number of chemical species that vary in relative strength. The absorbance spectrum is therefore a linear combination of individual absorbances with weight varying as a function of air mass and weather conditions (most notably from water). Note that single chemical species could contribute more than one independent absorbance spectrum as the specie is distributed vertically through the atmosphere (e.g. steam at 10$^\circ$C and at high altitude may contribute slightly different spectra with relative weights that are not necessarily correlated). So, if one can build a library of these individual absorbances, then a given science observation can be calibrated using a linear combination of these absorbances.

In practice, this strategy would be implemented by regularly observing telluric standards, ideally rapidly rotating hot stars (A and earlier), at a large variety of air masses and water column. This could be done during poor-seeing nights, a handful of times per night. From these observations, a large library of absorption spectra is built and a principal component analysis (PCA \footnote{ http://en.wikipedia.org/wiki/Principal\_component\_analysis }) is performed to identify independently varying absorbers. The PCA is done by expressing all observed absorption spectra in the library in absorbance. The multiplicative nature of absorption becomes a sum of various absorbances, which lends itself to least-square fitting. For every science spectra we build a least-square fit model with 1) an absorption-free spectra of a similar-type science target and 2) a linear combination of absorbance. This problem is very well constrained linearly as we will have $\sim$100\,000 resolution elements in the spectrum for only a few ($<10$) degrees of freedom. Whether this fit is performed piece-wise (e.g., one order at a time) or simultaneously through the entire wavelength domain remains to be seen. Note that very strong (i.e. $>$20\%) absorption features should not be masked at this point as they provide the best constraints on the relative weight of the PCA components in this fit. They can be rejected at a later time in the analysis, for example at the time of the RV measurement;

The advantages of this technique are important; one does not necessarily need simultaneous calibration as long as a large set of observations at similar air mass and water column are present in the input library, leading to an increase in observing efficiency. As the science data itself is used to derive the absorbance to be subtracted, so the calibration is perfectly simultaneous by nature. Furthermore, there is no need to explicitly add RV shifts in the absorption spectra, as the PCA library will, over time, include absorptions at velocities representative of what is expected for a given observation. The drifts in velocity will appear as PCA components. If only one species was present, the first PCA component would be that absorber's signature, while the second component would be the spectral derivative of flux as a function of wavelength. If the absorption has a non-zero velocity in a given observation, we will see changes in the amplitude of that second component. One could explicitly include an additional component when performing the absorption modeling by adding df/d$\lambda$. This has not been implemented yet but would require minimal effort, should the need arise.

One could wonder why such a technique has not been used in the past? The technique can only be used if the telluric spectra are properly resolved. Blends of numerous lines of differing strengths do not scale as the log of absorbance. Furthermore, the wavelength coverage is large enough so that the PCA reconstruction is well constrained. In practice, one needs a number of lines much larger than the number of independent degrees of freedom. In the optical, where most high-resolution spectrograph operate,  the interest is modest as most of the domain is relatively free of strong telluric lines. A few authors (Snellen et al.\cite{Snellen:2010}, Birkby et al. \cite{Birkby:2013}) have used a conceptually similar method, where they de-correlate telluric absorption from airmass. Here no telluric absorption model is used, and absorption depth is assumed to scale linearly with airmass, which may not be necessarily the case of chemical species, such as water, that have scale heights that differ significantly from that of the bulk of the atmosphere. This method has been applied in $K$ and $L$ band over limited domains. There has also been efforts to directly model telluric absorption in astronomical datasets, notably by the ESO team developping the $Molecit$\cite{Kausch:2014} toolkit\footnote{http://www.eso.org/sci/software/pipelines/skytools/molecfit}. Purely empiric methods such as the one proposed here and modelling from line lists are certainly not mutually exclusive, although hybrid methods have not yet been documented. One could, for example, imagine an initial removal of telluric absorption by modelling complemented by a PCA-based approach to the removal of residuals to that subtraction.


\section{Numerical implementation with HARPS datasets}

\begin{figure}
   \begin{center}
   \begin{tabular}{c}
   \includegraphics[height=7cm]{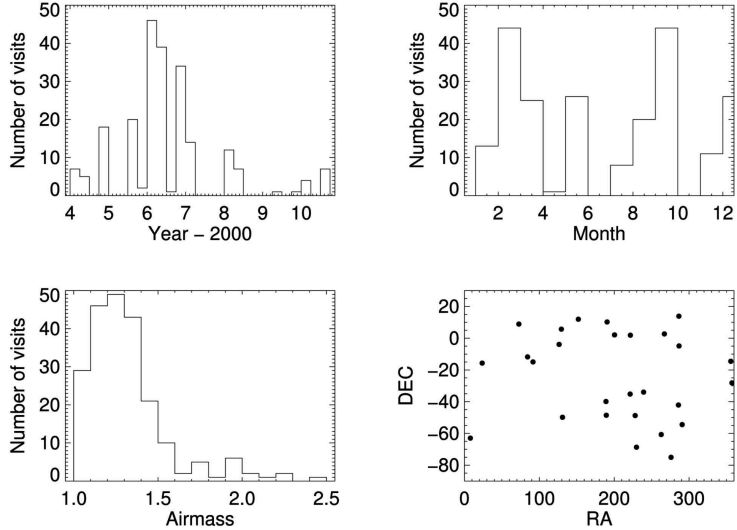}
   \end{tabular}
   \end{center}
   \caption[example] 
   { \label{fig2} Sky distribution of hot stars used to derive the telluric absorption models (low right). Air mass (lower left), monthly (upper right) and time (upper left) distribution of hot star observations.}
   \end{figure}

\begin{figure}
   \begin{center}
   \begin{tabular}{c}
   \includegraphics[height=7cm]{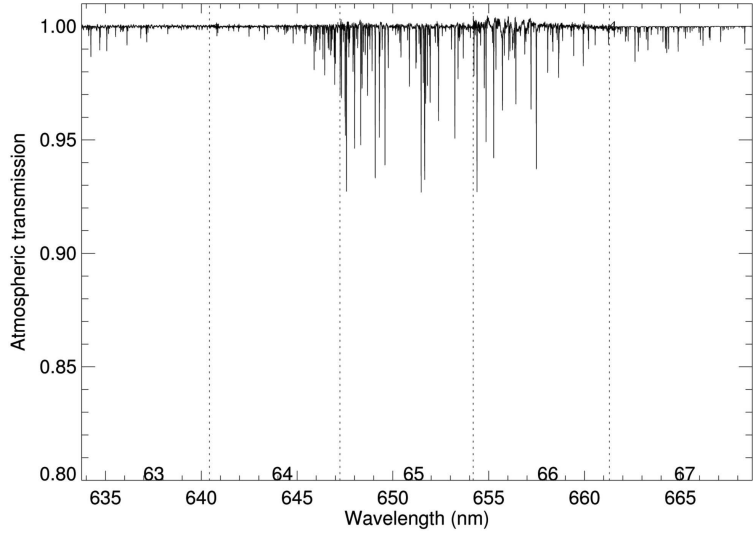}
   \end{tabular}
   \end{center}
   \caption[example] 
   { \label{fig3} Typical absorption spectrum for orders 63 through 67; numerous lines with a 1-8\% depth are present over this domain. This spectrum is derived from hot star observations. }
   \end{figure}

\begin{figure}
   \begin{center}
   \begin{tabular}{c}
   \includegraphics[height=7cm]{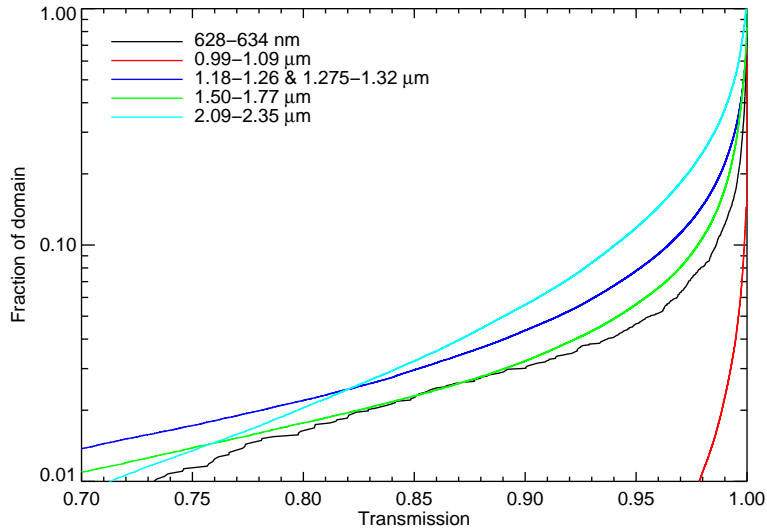}
   \end{tabular}
   \end{center}
   \caption[example] 
   { \label{fig4} Cumulative distribution of transmission within the bandpass shown in Figure\,\ref{fig3} and the four main near-IR windows. The $628-634$\,nm, $1.50-1.77\mu$m ($H$) and $2.09-2.35\mu$m ($K$) windows all have $\sim$2\% of their domains with transmission below 0.75. The $0.99-1.09\mu$m ($Y$) domain has a transmission below 98\% for only 1\% of the domain. Overall, the $628-634$\,nm has a telluric line content representative of what is seen in the nIR beyond $1.09\mu$m. The near-infrared absorption statistics is typical for the Mauna Kea ($\sim1.6$\,mm water column) and correspond to the statistics expected for SPIRou. The $628-634$\,nm data is for the HARPS dataset (La Silla) described here. Absorption for near-infrared bandpasses are taken from the TAPAS\cite{Bertaux:2014} atmosphere absorption models. }
   \end{figure}

In order to verify the interest of a PCA-based subtraction of telluric line, we use two datasets obtained with HARPS on $\tau$ Ceti (G8V) and GJ436 (M3.5V) having respectively  1526 and 169 observations. $\tau$ Ceti observations were averaged in 73 epochs to average-out stellar pulsations on timescales of $<15$\,min. GJ436 is known to host a super-Earth\cite{}, which radial velocity signal is clearly detected by the observations presented here. We also retrieved hot star spectra from the HARPS spectral library. A sample of 218 observations of 30 telluric standards with 3 to 23 visits per star were retrieved. The sky distribution of stars, airmass distribution, time distribution and yearly distribution of visits are shown in Figure~\ref{fig2}. Observations of the two RV stars and hot stars were not concomitant. The HARPS domain ($378-691$\,nm) covers telluric absorption features, especially around 630\,nm that are comparable in depth with the near-infrared domain. Figure\,\ref{fig3} illustrates the domain of interest ($628-634$\,nm or orders 63 to 67 in HARPS) for estimating the accuracy of the proposed technique in subtracting telluric absorption. As shown in Figure\,\ref{fig4}, absorption over this wavelength interval is comparable to the absorption seen in near-infrared bandpasses.

The spectra for all hot stars in the library were corrected for blaze function. Low-frequency slopes in the SED were corrected by fitting a low-order polynomial to the absorption-free domain and all orders were normalized to one. A PCA extraction was performed on all hot star spectra and PCA components were saved in decreasing order of importance. For the analysis here, we considered only the 5 first components. This choice was justified by the fact that components beyond the 5$^{\rm th}$ component visually show no line-like feature and that for orders where strong (i.e. $\>5$\%) absorption is present, the 5 first components accounted for $\>95$\% of the total variance, suggesting that no significant power is left in the remaining components.

For each spectrum, we then perform a least-square fit of the estimate of the star's spectrum (initially assumed to be 1 for all wavelength bins) and the first 5 PCA absorption components. This was performed on all observations, which provide a first estimate of the total absorption. This absorption is then subtracted for all observations, corrected spectra are then registered to a common barycentric velocity and median combined to produce a high S/N, cleaned, reference stellar spectrum. This spectrum, shifted to a given observation's barycentric velocity, is then used in the least-square fit of the PCA components for that observation. After this second iteration, a more accurate estimate of the telluric-free absorption spectrum could be derived and used in again to measure absorption more accurately. In practice, only two iterations were sufficient to obtain satisfactory a convergence at the level of the observational noise.

\begin{figure}
   \begin{center}
   \begin{tabular}{c}
   \includegraphics[height=7cm]{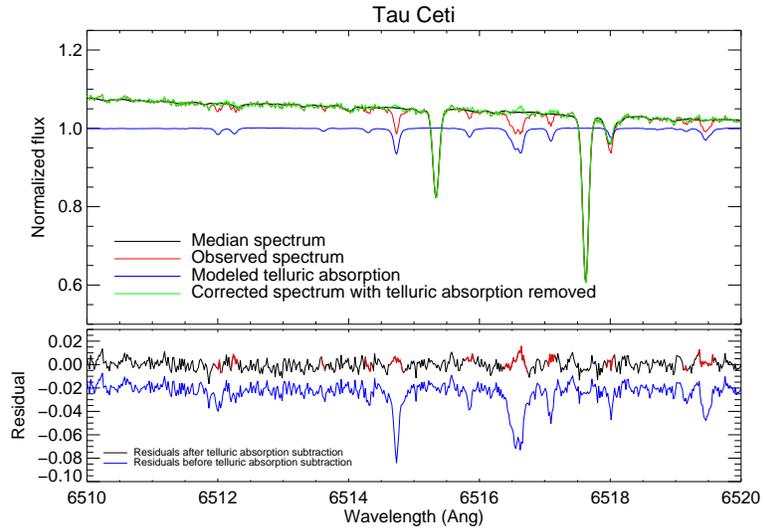}
   \end{tabular}
   \end{center}
   \caption[example] 
   { \label{fig5} Observed spectrum of $\tau$ Ceti over a wavelength domain affected by telluric absorption. This region is part of order 65 (see Figure\,\ref{fig3}).}
   \end{figure}

\begin{figure}
   \begin{center}
   \begin{tabular}{c}
   \includegraphics[height=7cm]{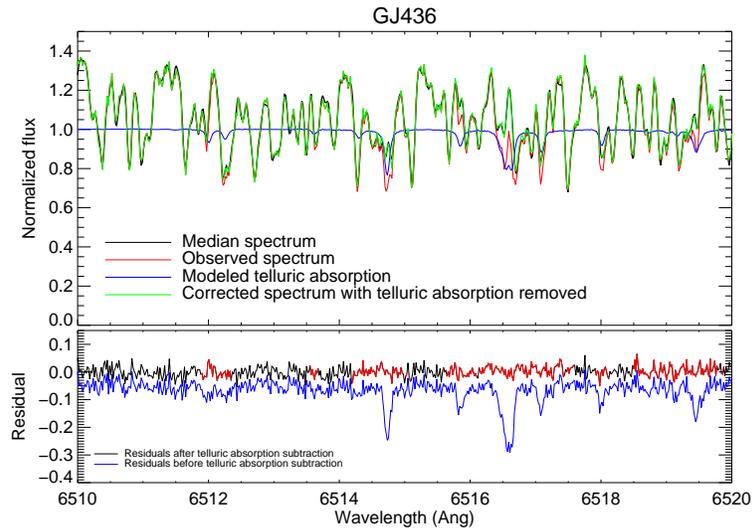}
   \end{tabular}
   \end{center}
   \caption[example] 
   { \label{fig6} Same as Figure\,\ref{fig5}, but for GJ436. The spectrum is dominated by molecular bands instead of atomic lines.}
   \end{figure}

\begin{figure}
   \begin{center}
   \begin{tabular}{c}
   \includegraphics[height=5cm]{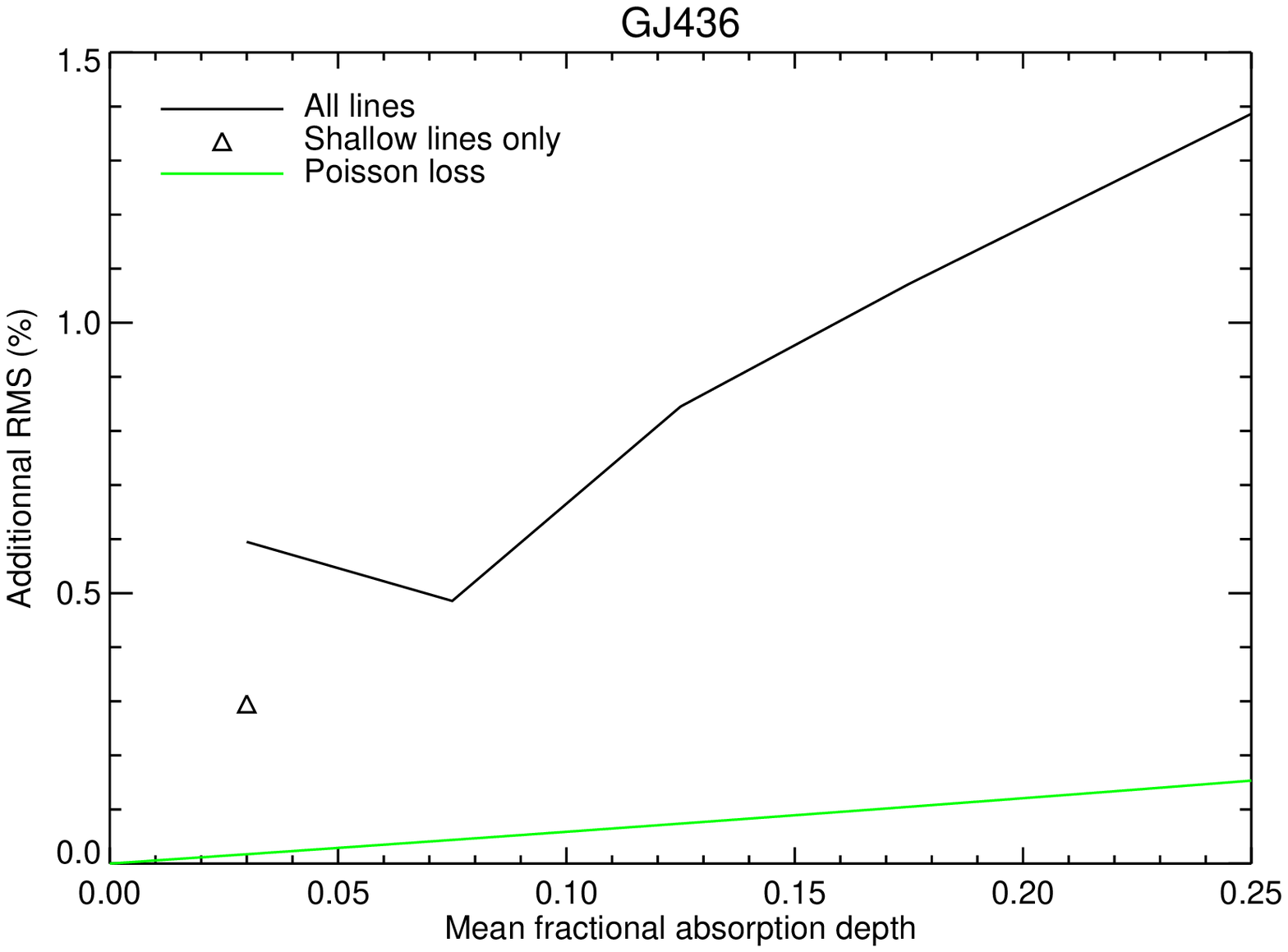}
   \includegraphics[height=5cm]{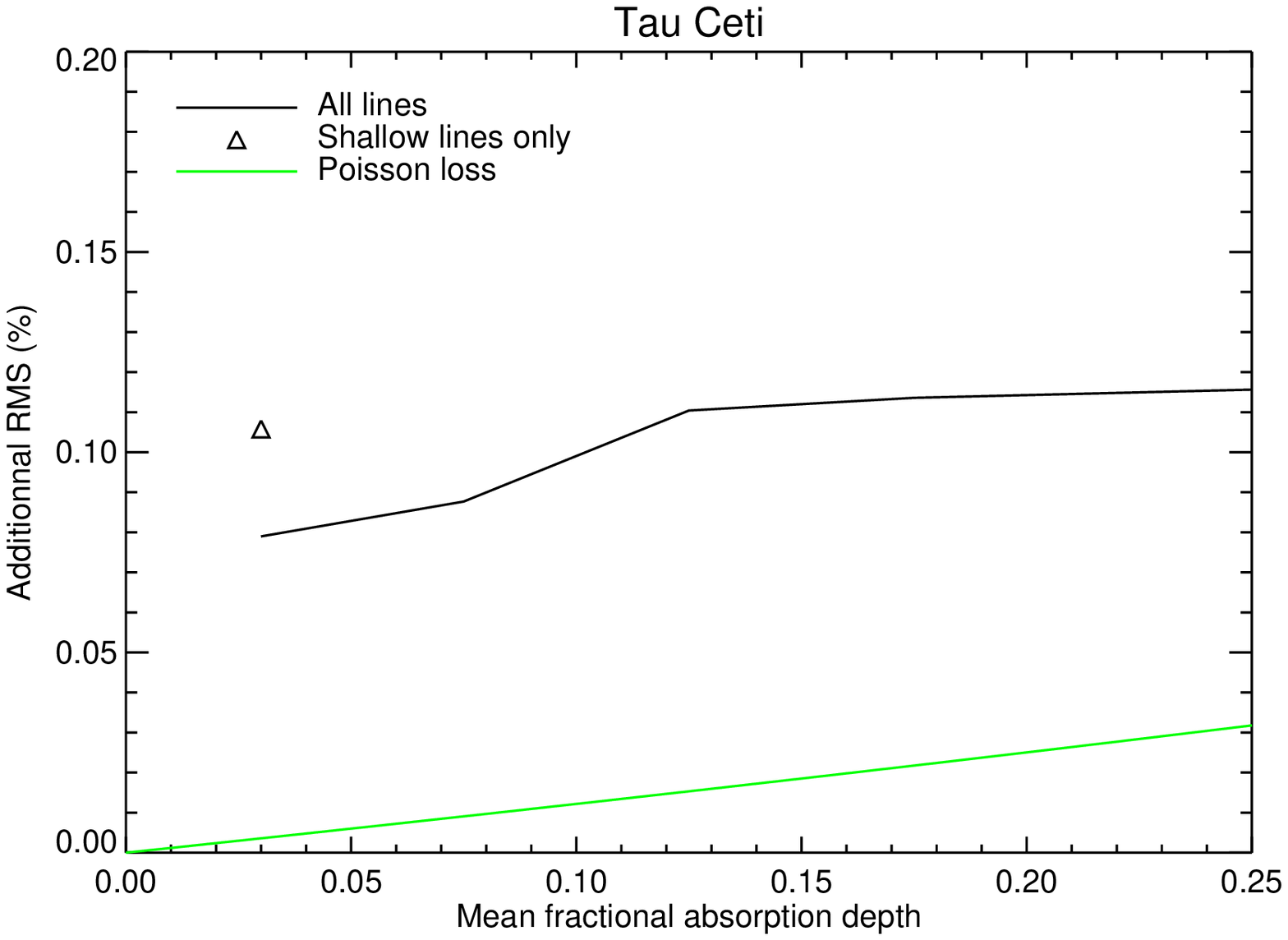}
   \end{tabular}
   \end{center}
   \caption[example] 
   { \label{fig7} Additional RMS from PCA subtraction as a function of absorption depth for GJ436 (left) and $\tau$ Ceti (right). In both cases, deeper absorption features lead to an increase in the residual RMS. For comparison, green lines show the relative increase in the Poisson noise from the loss in signal due to absorption. For $\tau$ Ceti, the increase in RMS due to telluric-line subtraction is about 0.1\% (right), while it is at the 1\% level from GJ436 (left). The “shallow lines only” represents the contribution of residuals for absorption between 1 and 5\% that is not adjacent to a deeper absorption (i.e. center of a shallow line in comparison of the wings of a deeper line).}
   \end{figure}

Figure\,\ref{fig5}  and Figure\,\ref{fig6}  respectively show the reconstructed telluric absorption for $\tau$ Ceti and GJ436 for sample science integration. The corrected spectra for a single observation are shown compared to the median spectra for the entire dataset shifted at the same barycentric velocity. Residuals between the fitted and observed spectra are shown. No obvious differences between regions affected by absorption and the rest of the domain is seen. Table\,\ref{tbl1} compiles the RMS of residuals for various levels of absorption. Figure\,\ref{fig7}  illustrates the same dataset and compares it with the additional RMS from the increase in the relative Poisson noise from decreased flux in areas affected by telluric absorption. Overall, the increase in RMS due to telluric subtraction is larger than can be accounted from only from increased Poisson noise due to the lower flux. The increase is also dependent upon the S/N of the observations; observations of $\tau$ Ceti show much lower RMS levels than GJ436, presumably from the poorer fit of telluric absorption, but this would need to be confirmed as GJ436 spectrum is dominated by molecular bands compared to atomic lines in $\tau$ Ceti. The spectrum of $\tau$ Ceti, dominated by atomic lines, may lend itself to better constraints on absorption.

\begin{table}[htbp]
\caption{Residual level (expressed in fractional RMS) between the cleaned spectra and reference spectrum as a function of telluric depth. For example, the residuals in the domain where there is $<1$\% absorption show a RMS of 1.15\% of the median flux level. RMS increases with telluric absorption. The contribution from subtraction has been estimated by quadratically subtracting the RMS without ($<1$\%) absorption from RMS observed at higher absorptions. For the $\tau$ Ceti, the additional RMS is at the level of 0.1\%, and $\sim1$\% for GJ436. Low ($1-5$\%) absorption regions have been divided in two categories; being either adjacent to deeper absorption or from isolated lines. For the GJ436 dataset, residuals are more than 2 times smaller with isolated shallow lines.} 
\label{tbl1}
\begin{center}       
\begin{tabular}{|c|c|c||c|c|} 

\hline
Absorption level & \multicolumn{2}{c||}{GJ436 (628-634 nm)} & \multicolumn{2}{c|}{ $\tau$ Ceti (628-634 nm) }\\
		 &  \multicolumn{1}{c}{$\sigma$} & \multicolumn{1}{c||}{	$(\sigma^2-\sigma^2_0)^{1/2} $}	&  \multicolumn{1}{c}{$\sigma$}	&  \multicolumn{1}{c|}{ $(\sigma^2-\sigma^2_0)^{1/2} $ }  \\

\hline\hline
$<1$\%			&1.15\%		&0.00\%		&0.24\%		&0.00\%\\
$1-5$\%			&1.19\%		&0.63\%		&0.26\%		&0.10\%\\
$1-5$\% not adjacent to deeper absorption	&1.31\%	& 0.29\%&	0.25\%&	0.09\%\\
$5-10$\%		&1.26\%		&0.52\%		&0.26\%		&0.09\%\\
$10-15$\%		&1.44\%		&0.86\%		&0.27\%		&0.12\%\\
$15-20$\%		&1.60\%		&1.11\%		&0.27\%		&0.12\%\\
$20-30$\%		&1.83\%		&1.43\%		&0.28\%		&0.14\%\\
\hline

\end{tabular}
\end{center}
\end{table}

\begin{table}[htbp]
\caption{Same as Table\,\ref{tbl1}, but for orders 62 through 68 ($634- 667$ nm)} 
\label{tbl2}
\begin{center}       
\begin{tabular}{|c|c|c||c|c|} 

\hline
Absorption level & \multicolumn{2}{c||}{GJ436 } & \multicolumn{2}{c|}{ $\tau$ Ceti  }\\
		 &  \multicolumn{1}{c}{$\sigma$} & \multicolumn{1}{c||}{	$(\sigma^2-\sigma^2_0)^{1/2} $}	&  \multicolumn{1}{c}{$\sigma$}	&  \multicolumn{1}{c|}{ $(\sigma^2-\sigma^2_0)^{1/2} $ }  \\

\hline\hline
$<1$\%&	0.97\%&	0.00\%&	0.23\%&	0.00\%\\
$1-5$\%&	1.03\%&	0.33\%&	0.28\%	&0.15\%\\
$1-5$\% not adjacent to deeper absorption&	1.04\%&	0.36\%&	0.29\%&	0.17\%\\
$5-10$\%&	1.06\%&	0.43\%&	0.32\%&	0.22\%\\
\hline

\end{tabular}
\end{center}
\end{table}

Overall, a PCA-based analysis shows that one can subtract telluric features to better than a factor of 10 using non-contemporaneous telluric absorption measurements. This technique is likely to be key for near-IR velocimeters in order to maximize the useful RV domain. The technique is of general interest for ground-based optical and near-infrared observations and is by no means limited to RV measurement. It could open-up significant parts of the near infrared to spectroscopic observation, such as the wavelength interval between $H$ and $K$ under low water vapor conditions. Another important avenue to explore in the near future is the use of synthetic spectra (e.g. the TAPAS library\footnote{http://www.pole-ether.fr/tapas/}) to derive PCA components. Assessing the interrest and the limitations of using modeled spectra of the atmosphere to subtract absorption is the logical next step in this work.

\section{Telluric line subtraction and RV accuracy}
The PCA telluric absorption subtraction has been performed on GJ436 and $\tau$ Ceti HARPS datasets. This allows one to verify that the method does not induce spurious RV signal and, hopefully, slightly improve the HARPS RV measurements that may suffer from low-level absorption unaccounted for by its current pipeline. Overall, there is always a gain in RV accuracy from the use of PCA subtraction (see Table\,\ref{tbl3}). It is important to note that GJ436 has significant RV jitter that is not accounted for by photon noise and instrumental effects alone. This could be due to additional planets in its system and/or stellar activity.

The $\tau$ Ceti dataset has an $\sigma_{\rm o-c}$ of $1.07\pm0.09$\,m/s with the original spectrum and $0.93\pm0.08$\,m/s with PCA subtracted spectrum, clearly demonstrating that the technique described here does not degrade RV accuracy, but that it also brings an improvement that could be due to the correction of absorption lines that were not rejected by the HARPS pipeline\footnote{http://www.eso.org/sci/facilities/lasilla/instruments/harps/doc/DRS.pdf}. This results is confirmed by the fact that the measurement performed only on orders 63 through 67 (See Figure\,\ref{fig3}) without excluding the many 1-8\%-deep lines gives a 10.3 m/s RMS with the original spectra (implying that telluric lines completely dominate the RV RMS error budget) and a 1.58\,m/s RMS with the PCA method. The fact that we obtain an RV accuracy of $\sim1.6$\,m/s over a wavelength domain with a large number of telluric lines and $<1$\,m/s over the entire HARPS domain shows that the PCA method can be successfully applied to obtain m/s RV precision and that no show-stopper is present to would prevent us from applying this method to the nIR. Furthermore, as quantified in Table\,\ref{tbl4}, there is a significant gain in spectral domain from telluric line subtraction. Over orders $63-67$, $\sim18$\,\% of the domain is lost to tellurics before subtraction, and this value falls to $\sim5$\,\% after subtraction. 


\begin{table}[htbp]
\caption{$\sigma_{\rm o-c}$ for GJ436 and $\tau$ Ceti RV data with and without telluric line subtraction. In all cases, the PCA telluric line subtraction improves the RV measurements. Measurements performed on orders 63-67 sample parts of the $R$ band with significant telluric absorption. } 
\label{tbl3}
\begin{center}       
\begin{tabular}{ |p{8cm}|c|c|p{3cm}|} 
\hline
&GJ436&$\tau$ Ceti&Rejected domain due to tellurics\\
\hline\hline
Original spectrum, excluding strong telluric absorption not well removed in RV measurement & 1.43\,m/s & 1.07\,m/s & 4\% \\
\hline
PCA subtracted spectrum, excluding telluric absorption in RV measurement & 1.39\,m/s & 0.93\,m/s & 1.1\% \\
\hline
Original spectrum, orders $63-67$, excluding telluric absorption in RV measurement & 2.30\,m/s & 3.6\,m/s & 17.7\% \\
\hline
PCA subtracted spectrum, Orders $63-67$, excluding strong telluric absorption in RV measurement & 2.21\,m/s & 1.77\,m/s & 4.9\% \\
\hline
PCA subtracted spectrum, orders $63-67$, including regions where strong telluric absorption has been subtracted	&2.64\,m/s&	1.58\,m/s&\\
\hline 
\end{tabular}
\end{center}
\end{table}

   \begin{figure}
   \begin{center}
   \begin{tabular}{c}
   \includegraphics[height=4cm]{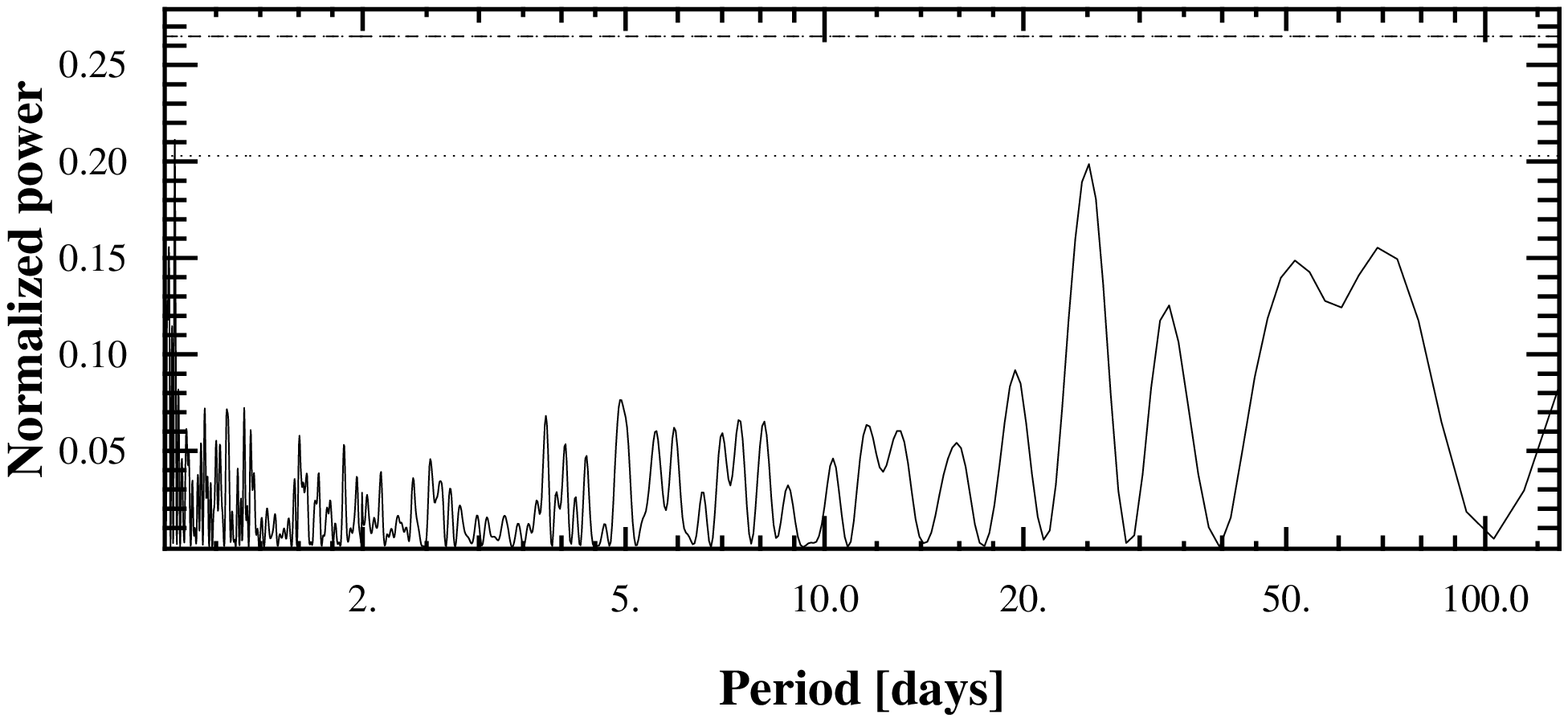}
   \includegraphics[height=4cm]{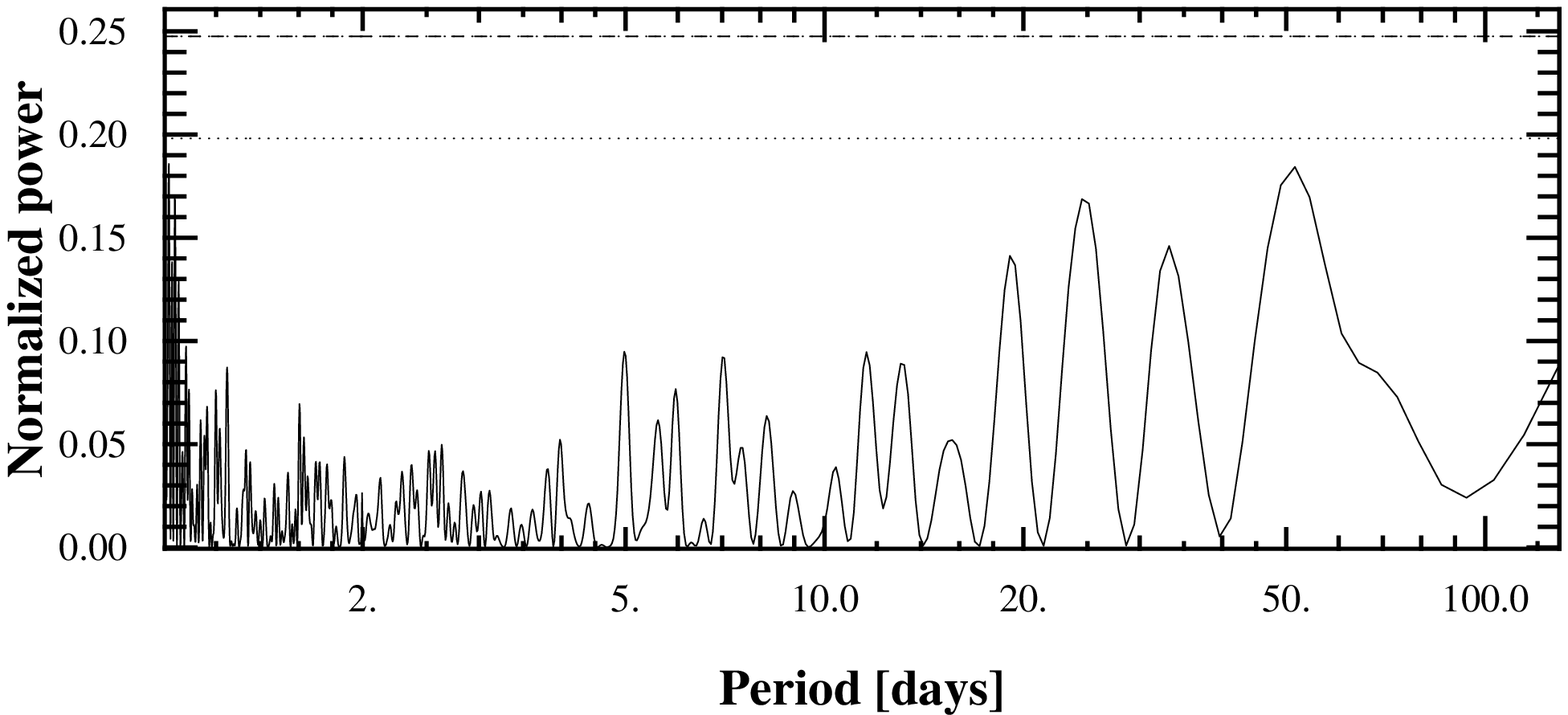}
   \end{tabular}
   \end{center}
   \caption[example] 
   { \label{fig9} $\tau$ Ceti periodogram with (left) and without (right) PCA cleaning of telluric absorption. No spurious periodogram peak is added from the PCA subtraction. Strong telluric residuals should leave a periodogram peak at 1 year and its harmonics, but the observations spawn a baseline too short to show this peak.}
   \end{figure} 

\section{The benefit of variable telluric absorption mask}
The PCA subtraction of telluric lines as described here provides a mean of efficiently subtracting telluric lines. It is not known yet to which depth lines can be accurately subtracted and still allow for m/s velocimetry. We demonstrate that lines as deep as $\sim$10\% are manageable, but what about deeper lines? The HARPS dataset does not allow one answer this question and, at a certain level depth of absorption, one will need to mask telluric lines and exclude the affected domain from the analysis. 

OWhen few lines are present, one can exclude domain within $\pm30$\,km/s of strong absorption lines so that the excluded domain remains constant through the year as the line-of-sight barycentric velocity changes. This increases by a factor $\sim10$ the domain lost to absorption (telluric lines are a few km/s wide but this requires excluding $\sim60$\,km/s windows), but ensures a measurement over the same spectroscopic features through the year (see Table\,\ref{tbl4}). Absorption depths were determined from the ATRAN  atmosphere model and a 1.5\,mm (i.e. Mauna Kea median) water column. Without a rejection of a $\pm30$\,km/s interval, the loss in RV signal is moderate (3-15\%) for the $J$, $H$ and $K$ bands, even with relatively stringent rejection criteria ($<$5\%). This loss increases rapidly ($\sim$20 to $\sim$60\%) when rejecting $\pm30$\,km/s around telluric lines. 

\begin{figure}
   \begin{center}
   \begin{tabular}{c}
   \includegraphics[height=7cm]{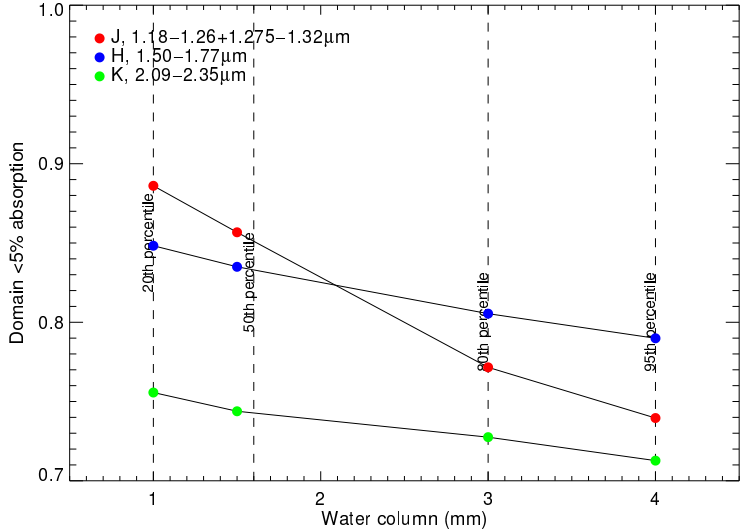}
   \end{tabular}
   \end{center}
   \caption[example] 
   { \label{fig10} Fraction of the wavelength domain lost as a function of water . In practice, the difference between the 20$^{\rm th}$ and 95$^{\rm th}$ percentile for the Mauna Kea is relatively modest. This is due to the presence of saturated lines that vary little with water column.}
   \end{figure}

By using a mask that only removes telluric lines of a given epoch, one can significantly increase the usable domain. In the optical, the overall gain is minimal, but in the nIR, this could lead to a significant increase in the RV signal. We also show (Figure\,\ref{fig10}) that there is little changed in the domain lost to telluric absorption with the water column, suggesting that RV observations can be performed on most ($>95$\,\%) of the nights.

\begin{table}[htbp]
\caption{Loss in domain from telluric absorption with and without rejection of a $\pm30$\,km/s interval surrounding absorption lines.} 
\label{tbl4}
\begin{center}       
\begin{tabular}{|c|c|c|c|} 

\hline
Absorption level & \multicolumn{3}{c|}{ Domain lost }\\
		 & \multicolumn{1}{c}{ 1.18-1.26 \& 1.275-1.32$\mu$m } &  \multicolumn{1}{c}{ 1.50-1.77$\mu$m }&  \multicolumn{1}{c|}{ 2.09-2.35$\mu$m }  \\

\hline

$<5$\%			& 10.7\%	& $9.2$\%	& $22.9$\%\\
$<20$\%			& 2.8\%		& $3.3$\%	& $6.8$\%\\
$<5$\%, $\pm30$\,km/s	& 40.9\%	& $37.0$\%	& $57.0$\%\\
$<20$\%, $\pm30$\,km/s	& 14.3\%	& $16.5$\%	& $25.6$\%\\

\hline

\end{tabular}
\end{center}
\end{table}

\subsection{Radial velocity measurements with 10\% and 20\% masked domain}
\begin{figure}
   \begin{center}
   \begin{tabular}{c}
   \includegraphics[height=7cm]{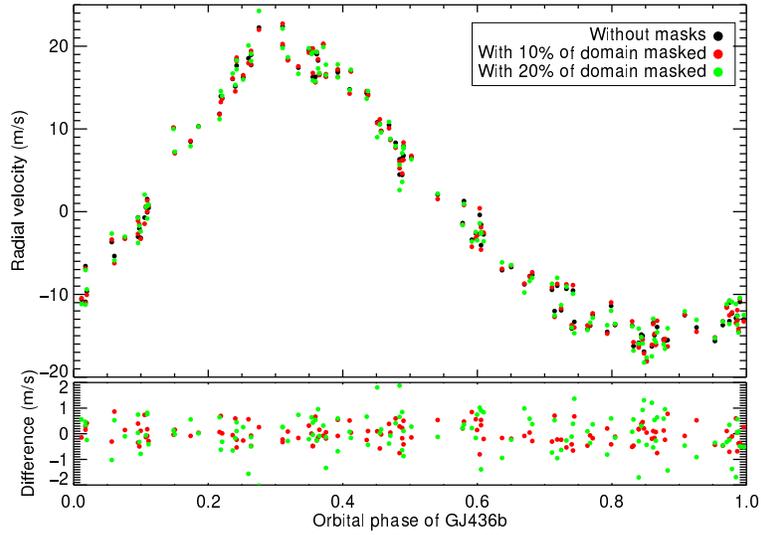}
   \end{tabular}
   \end{center}
   \caption[example]
   { \label{fig11}Radial velocity measured with the entire HARPS spectrum (black) and with 10\% of the domain excluded with 8 km/s-wide windows (red) and with 20\% of the domain excluded (green). The RV measurements closely match; differences between the two measurements have an RMS of $0.37/0.75$ (red/green) m/s and show now specific structure as a function of phase.
  }
   \end{figure}

\begin{figure}
   \begin{center}
   \begin{tabular}{c}
   \includegraphics[height=7cm]{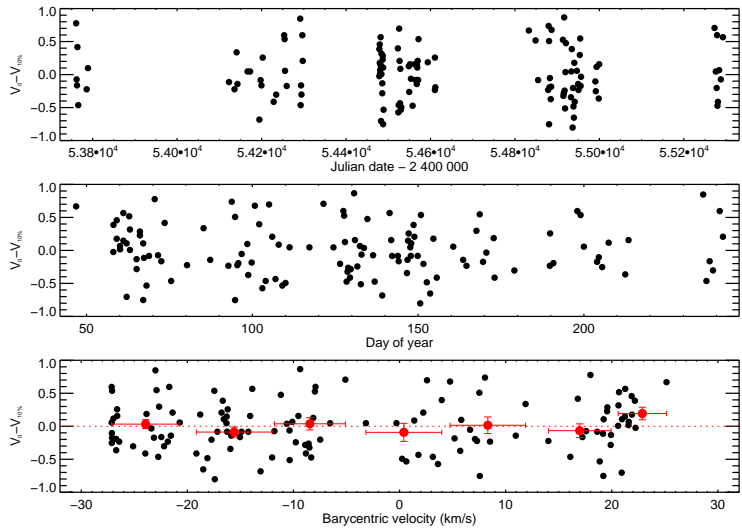}
   \end{tabular}
   \end{center}
   \caption[example] 
   { \label{fig12}RV difference between velocities measured with and without 10\% mask for GJ436. Panels respectively show the differences as a function of date, day of year and barycentric velocity. No systematic trend can be seen with respect to date of barycentric velocity, suggesting that no systematic effect at the tens of cm/s is present and that the RMS is dominated by the increase in photon noise. In the lower panel, red symbols show the mean velocity difference per 8 km/s bin; error bars correspond to the standard error for each of these means.  }
   \end{figure}

We performed an analysis of the GJ436 and $\tau$ Ceti HARPS RV data to determine whether having a fixed mask on telluric lines that is not tracking the barycentric velocity of the star would induce significant systematic effects. We defined a mask consisting of 8\,km/s-wide gaps randomly distributed across the HARPS domain and with a fractional coverage of 10\% (roughly what is expected in $H$ and $K$) and 20\% (more than any nIR photometric bandpass). The radial velocity was extracted from the dataset with and without masking. From photon-counting statistics alone, one would expect a 5\% et 10\% decrease in the RV accuracy. Figure\,\ref{fig11} shows the RV measurements (black, all wavelength domain, red, excluding 10\%, green, excluding 20\%) phased to GJ436's planet orbital period (2.643 days). The rejection of $10\%$ of the wavelength domain to simulate the presence of numerous absorption lines only increases the RV RMS by 0.38 m/s (see Table\,\ref{tbl5}). Figure\,\ref{fig12} shows RV differences as a function of time, day of year and barycentric velocity. No obvious correlation exists between RV measurement and either barycentric velocity or day of year.

A similar analysis has been performed on $\tau$ Ceti's data and the impact of masking 10\% and 20\% of the domain is minimal ($<10$\,cm/s) and within the uncertainties on the RMS. This strongly suggests that removing 10\% of the domain introduces no systematic bias and that there is no reason to take the conservative approach to reject a $\pm30$\,km/s on either side of telluric lines.

In order to establish the level of systematic residuals from masking, we binned velocity differences as a function of barycentric velocity (8\,km/s-bins, see red symbols in Figure\,\ref{fig11}). The $\chi$2  ($\chi 2=4.6$ for 7 degrees of freedom, p=0.29) for these measurements is consistent with no systematic correlated with the barycentric velocity. Statistically, upper limits can be set on any signal correlated with barycentric velocity from numerical simulations. We set a 1-$\sigma$ and 2-$\sigma$ upper limits of 0.13 and 0.24\,m/s on any correlated signal between barycentric velocity and difference in RV. The usage of a mask that varies with time significantly improves the RV budget and is not mutually exclusive with the PCA rejection. It would be desirable to both subtract telluric lines and use a rejection mask, the depth of which needs to be determined with actual data.

\begin{table}[htbp]
\caption{Dispersion in (O-C) RV measurements with the full HARPS coverage ($\sigma_0$), the rejection of 10\% of the domain to simulate numerous telluric lines ($\sigma_{10\%}$) and the difference between the two RV measurements ($\sigma_{diff}$).} 
\label{tbl5}
\begin{center}       
\begin{tabular}{|c|c|c|} 

\hline
	&GJ436	&$\tau$ Ceti\\
\hline
\hline
$\sigma_0$ (RV measurements)&	1.39 m/s&	0.93 m/s\\
\hline
$\sigma_{10\%}$ (RV measurements excluding 10\% of domain)&	1.46 m/s&	0.86 m/s\\
\hline
$\sigma_{20\%}$  (RV measurements excluding 20\% of domain)&	1.60 m/s&	0.94 m/s\\
\hline
$\sigma_{diff}$ (no masking vs 10\% masking)&	0.38 m/s&	0.16 m/s\\
\hline
$\sigma_{diff}$ (no masking vs 20\% masking)&	0.75 m/s&	0.20 m/s\\

\hline

\end{tabular}
\end{center}
\end{table}


\bibliography{bibdesk}
\bibliographystyle{spiebib}   

\end{document}